# Preferential evaporation in atom probe tomography: an analytical approach

*Constantinos Hatzoglou[1,2], Solène Rouland[1], Bertrand Radiguet[1], Auriane Etienne[1], Gérald Da Costa[1], Xavier Sauvage[1], Philippe Pareige[1], François Vurpillot[1]*

[1]Normandie Université, UNIROUEN, INSA Rouen, CNRS, Groupe de Physique des Matériaux, 76000 Rouen, France

[2]Department of Materials Science and Engineering, NTNU, Norwegian University of Science and Technology, Trondheim 7491, Norway

**Abstract**

Atom probe tomography (APT) analysis conditions play a major role in the composition measurement accuracy. Preferential evaporation, which significantly biases apparent composition, more than other well-known phenomena in APT, is strongly connected to those analysis conditions. One way to optimize them, in order to have the most accurate measurement, is therefore to be able to predict and then to estimate their influence on the apparent composition. An analytical model is proposed to quantify the preferential evaporation. This model is applied to three different alloys: NiCu, FeCrNi and FeCu. The model explains not only the analysis temperature dependence, as in already existing model, but also the dependence to the pulse fraction and the pulse frequency. Moreover, the model can also provide energetic constant directly linked to energy barrier, required to field evaporate atom from the sample surface.

## 1. Introduction

Atom probe tomography (APT) is a three-dimensional (3D) characterization technique that is nowadays commonly used to visualize and quantify the microstructure of materials at atomic scale. This technique is widely used in materials sciences. More details on this technique can be found in (Lefebvre et al., 2016; Baptiste Gault et al., 2012; Miller, 2000).

APT is based on the field evaporation of atoms located at the surface of the specimen. The elemental nature of each evaporated atom is obtained by time-of-flight mass spectrometry. Such measurement requires a precise control of the field evaporation (FE) process. This is done by superimposing to the constant voltage ($V_{DC}$) applied to the specimen, short pulses that trigger the field evaporation and, at the same time, these pulses trigger time of flight measurements. If some atoms are FE between pulses, their time of flight is unknown, and they are not considered for the composition quantification. This phenomenon is called preferential evaporation (PE) (Tsong, 2005; Miller, 2000; Miller et al., 1996; Miller & Forbes, 2009) when this bias is element specific. PE may degrade significantly the composition measurement accuracy in APT.

The probability for an atom to be FE is linked to many parameters such as the temperature of the sample, the electric field at the sample surface and the evaporation field of the considered atoms. Evaporation field is defined as the required field for an atom to be evaporated from the specimen surface (Tsong, 2005). At defined conditions of temperature and electrical field, if atoms of a compound have similar evaporation field then their probability to be FE at $V_{DC}$ is similar. In this case the global number of atoms evaporated out of pulses follows the actual composition of the specimen, and the composition measurement made from atoms collected on pulses is not biased. However, if the evaporation field between atoms is significantly different, some species exhibit a higher probability to be FE out of pulses. This generates some deficit of these species and the composition measurement can be significantly biased. Indeed, since the evaporation probability is not the same out of pulses, it affects then the residual proportion of atoms that can be evaporated on pulses.

These biases have been already observed and studied by many authors in different materials such as FeCu (Takahashi et al., 2017), cemented tungsten carbide (Peng et al., 2017), GaAs (Russo et al., 2017), ODS steel (Hatzoglou et al., 2017) or ZnO (Amirifar et al., 2015). It has been determined that the measured elemental composition depends on the APT analysis conditions. In some cases, for example ZnO or GaAs, the situation is even more complex due to molecular ion dissociation (Blum et al., 2016; Zanuttini et al., 2017; Saxey, 2011) or because of detection limitations like for tungsten carbide (Bacchi et al., 2019; Meisenkothen et al., 2015).

It is generally accepted that quantitative analyses are preferably obtained under low specimen temperature, high-pulse fraction (ratio of the pulses amplitude over the DC voltage) and high pulse repetition rate. Unfortunately, these optimal conditions also increase the sample failure probability (Prosa et al., 2019). A solution would be to develop an analytical model to better predict such phenomena. This model would provide the more favourable analysis conditions while maintaining high measurement accuracy and low bias. A first model was proposed by Takahashi *et al.* (Takahashi & Kawakami, 2014). This model explains the sample temperature dependence on the apparent composition of solute element, whereas the dependence on the pulse fraction and pulse frequency is not completely predicted.

In this study, a new analytical model of the PE for an electrical pulse APT is proposed. This model was applied to the experimental biases observed in three model alloys studied at different APT analysis conditions. More than reproducing the influence of all the studied parameters (temperature, pulse fraction and repetition rate), the model can also provide some valuable information about energetic constant directly linked to energy barrier, required to FE atom from the sample surface.

**2. The analytical model**

In APT, several biases may arise as low FE occurs at the DC voltage (*i.e.* between two pulses). For a given material, the number of atoms FE between pulses may be estimated with respect to the number of atoms evaporated on pulses. In voltage mode APT, atoms evaporated on pulses have higher kinetic energy, and are identified on the mass spectrum. Atoms FE between pulses and detected without any correlation with pulses, are identified as background noise events (Lefebvre et al., 2016).

For an alloy composed of $j$ chemical species, the number of atoms $i$ (among the $j$ elements) FE on pulses, $N_{P,i}$ can be calculated, regarding its own evaporation pulse width $\tau_{p,i}$ (ns range) and assuming that its rate of evaporation ($K_{p,i}$) is defined by an Arrhenius expression,:

$$N_{p,i} = \tau_{p,i} K_{p,i} = \tau_{p,i} \nu N_i \times exp\left(-\frac{Q_i(F_{T,i})}{k_B T}\right) \quad (1)$$

with $K_{p,i}$ the evaporation rate of atoms $i$ during the pulse, $\nu$ the surface atoms vibration frequency, $N_i$ the total number of $i$ atoms FE per unit of time, $k_B$ the Boltzmann constant, $T$ the analysis temperature, and $Q_i$ the energy barrier for an atom $i$ at a defined field $F_{T,i}$. $F_{T,i}$ is the total electric field above atoms $i$ at the pulse maximum:

$$F_{T,i} = F_{p,i} + F_{DC,i} \quad (2)$$

with $F_{DC,i}$ the electric field at the DC voltage and $F_{p,i}$ the pulse electric field.

Close to the evaporation field, the expression of the energy barrier is given by the common expression:

$$Q_i(F_{T,i}) = C_i \left(1 - \frac{F_{T,i}}{F_{EV,i}}\right) \quad (3)$$

with $C_i$ an energetic constant of the atoms $i$ and $F_{EV,i}$ its evaporation field (Lefebvre et al., 2016; Forbes, 1995; Miller & Forbes, 2009).

The agreement of this last equation with experimental measurement is correct, if we assume an energetic constant $C_i$ in the range 1-3 eV for refractory metals (Kellogg, 1984; Ernst, 1979; Wada, 1984). Experimentally, the linearity of $Q_i(F)$ is approximately observed in the range of $0.8 F_{EV,i}$ to $0.95 F_{EV,i}$ for some materials (Kellogg, 1984). Deviations exist out of this region but there will be neglected in a first approach. The energetic constant $C_i$ is a projection of $Q_i(F_i)$ at a field close to the evaporation field $F_{EV,i}$ according to the linear approximation of $Q(F)$.

In this range, the linear tendency has also been observed with ab initio calculation by Sanchez *et al.* (Sánchez et al., 2004). Note that the experimental estimation of $C_i$ equal to $Q_{0,i}/2$ predicted by the Muller-Schottky model is poor as discussed by different authors (with $Q_{0,i}$ the energy barrier without field in the Muller-Schottky description of the field evaporation)(Muller et al., 1965).

The number of atoms $i$ evaporated on one pulse is then:

$$N_{p,i} = \tau_{p,i} \nu N_i \times exp\left(-\frac{C_i}{k_B T}\left(1 - \frac{F_{T,i}}{F_{EV,i}}\right)\right) \quad (4)$$

Similarly, the number of atoms $i$ evaporated between two pulses is:

$$N_{DC,i} = \tau_{DC,i} K_{DC,i} = \tau_{DC,i} \nu N_i \times exp\left(-\frac{C_i}{k_B T}\left(1 - (1 - \alpha_p)\frac{F_{T,i}}{F_{EV,i}}\right)\right) \quad (5)$$

with $\tau_{DC,i}$ the time interval between two pulses (10 µs at 100 kHz) and $\alpha_p$ the ratio between the total and pulse electric field (i.e. $\alpha_p = F_p/F_T$). This ratio is related to the pulse fraction ($f_p$), commonly used experimentally. The pulse fraction is the ratio between the pulsed electric field and the field at the DC voltage (i.e. $f_p = F_p/F_{Dc}$) and is so linked to the ratio $\alpha_p$ by:

$$\alpha_p = \frac{f_p}{1 + f_p} \quad (6)$$

The apparent composition measured by APT could be estimated by writing a conservation law that assumes that the number of evaporated atoms over a complete cycle of evaporation, $N_C$ (DC + pulse) is constant. Considering the initial composition of atoms $i$, $X_i$, ($X_i = N_{C,i}/N_C$ with $N_{C,i}$ the number of atoms $i$ evaporated over a complete cycle of evaporation), and the equations (4) and (5), this conservation law can then be written as:

$$\frac{1}{X_i}(\tau_{p,i} K_{p,i} + \tau_{DC,i} K_{DC,i}) = \gamma \quad (7)$$

with $\gamma$ a constant.

It is important to stress that the electric fields above atoms at $F_{T,i}$ are not identical according to their chemical nature. In fact, due to the steep nature of the evaporation rate with the electric field, at the top of the voltage pulse, the electric field is very close to the evaporation field of each element. It is thus introduced $F_{T,i} = (1 - \varepsilon_i)F_{EV,i}$ with $\varepsilon_i$ a small value ($\varepsilon_i \ll 1$). It will be then considered $\varepsilon_i$ as negligible and so that $F_{T,i} = F_{EV,i}$. It is thus possible to correlate the evaporation rate at the DC voltage ($K_{DC,i}$) and the one at the pulse ($K_{p,i}$).

$$K_{DC,i} = K_{p,i} \times exp\left(-\frac{C_i}{k_B T}\alpha_p\right) \quad (8)$$

By combining this last equation and the equation (7), the evaporation rate on the pulse ($K_{p,i}$) is thus equal to:

$$K_{p,i} = \frac{\gamma}{\tau_{p,i}} \times \frac{X_i}{1 + \frac{\tau_{DC,i}}{\tau_{p,i}} exp\left(-\frac{C_i}{k_B T}\alpha_p\right)} \quad (9)$$

Considering a gaussian shape electrical pulse, (Fig. 1), with a FWHM Δ, the FE pulse FWHM (i.e. $\tau_{p,i}$) is expressed as follows:

$$\tau_{p,i} = \frac{\Delta}{2}\sqrt{-\frac{1}{ln(2)} \times ln\left(1 - ln(2)\frac{k_b T}{C_i \alpha_p}\right)} \quad (10)$$

This equation is obtained from the evaporation rate ($K_N$, Fig. 1), equal to $exp(C(1 - F/F_T)/k_b T)$, with a temporal evolution of the normalized field ($F/F_T$) equal to $(exp(-t^2/2\sigma^2)F_p + F_{DC})/F_T$, with $\sigma = \Delta/2\sqrt{2ln2}$ regarding the gaussian pulse shape. In this study the pulses width ($\Delta$) is 1 ns since a LEAP 4000 HR has been used. The FE pulse width is temporally much narrower than the electrical pulse one (Fig. 1).

As a function of their energetic constant ($C_i$), the evaporation pulse duration seen by the atoms is not the same (Fig. 1). This equation restricts the study to a certain range of analysis conditions, since $ln(2) k_b T/C_i \alpha_p$ must be lower than 1. However, as it will be demonstrated later, this limitation range is very far from the commonly used experimental conditions. The ratio $\tau_{DC,i}/\tau_{p,i}$ can be expressed as a function of APT parameters and is then equal to $(1 - f\tau_{p,i})/f\tau_{p,i}$ with $f$ the pulse frequency and $\Delta$ the pulse width.

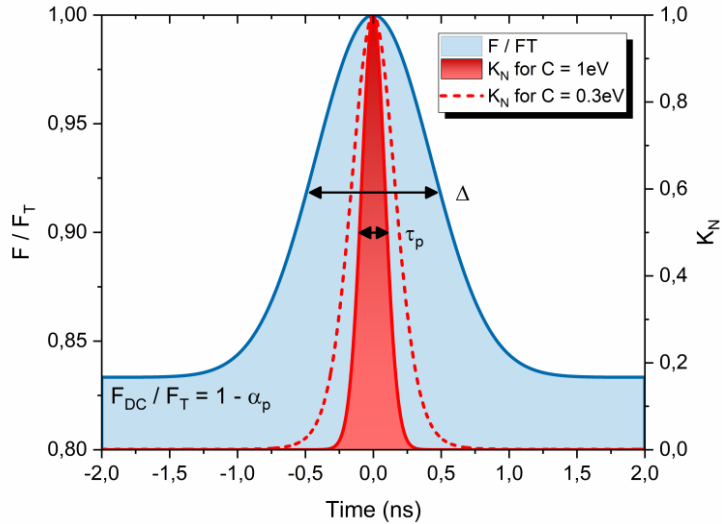

**Fig. 1**: Temporal (ns) evolution of the normalized field ($F/F_T$) at the sample apex during a gaussian shape electric pulse (FWHM $\Delta$ = 1 ns) and its associated evaporation rate ($K$). The evaporation rate has been plotted for two different energetic constants ($C$): 1 (red line) and 0.3 eV (red dash line), at a temperature of 80 K and a pulse fraction of 20 %.

The apparent composition in atoms $i$ measured in APT ($X'_i$) is the ratio between the number of those atoms ($N_{p,i}$) and the total number of atoms both evaporated during the pulses. Considering, the expression of those numbers (equation 1), the apparent composition $X'_i$ can be written as:

$$X'_i = \left(1 + \sum_{k=1, k \neq i}^{j} \frac{\tau_{p,k} K_{p,k}}{\tau_{p,i} K_{p,i}}\right)^{-1} \quad (11)$$

Considering the expression of the evaporation rate of the atoms $i$ at the pulse $K_{p,i}$ (expression 9), the apparent composition in atoms $i$ measured in APT eventually becomes:

$$X'_i = \left(1 + \sum_{k=1, k \neq i}^{j} \frac{\tau_{p,k} X_k}{\tau_{p,i} X_i} \times \frac{f\tau_{p,i} + (1 - f\tau_{p,i}) exp\left(-\frac{C_i}{k_b T} \alpha_p\right)}{f\tau_{p,k} + (1 - f\tau_{p,k}) exp\left(-\frac{C_k}{k_b T} \alpha_p\right)}\right)^{-1} \quad (12)$$

The influence of the analysis conditions ($\alpha_p$, $T$ and $f$) on the apparent composition ($X'_i$) is not obvious, due to the complexity of equation (12). For this purpose, the evolution of the apparent composition for an equiatomic alloy (A/B, *i.e.* $X_A = X_B = 50$ at.%) was computed as a function of the different parameters of equation (12) (Fig. 2).

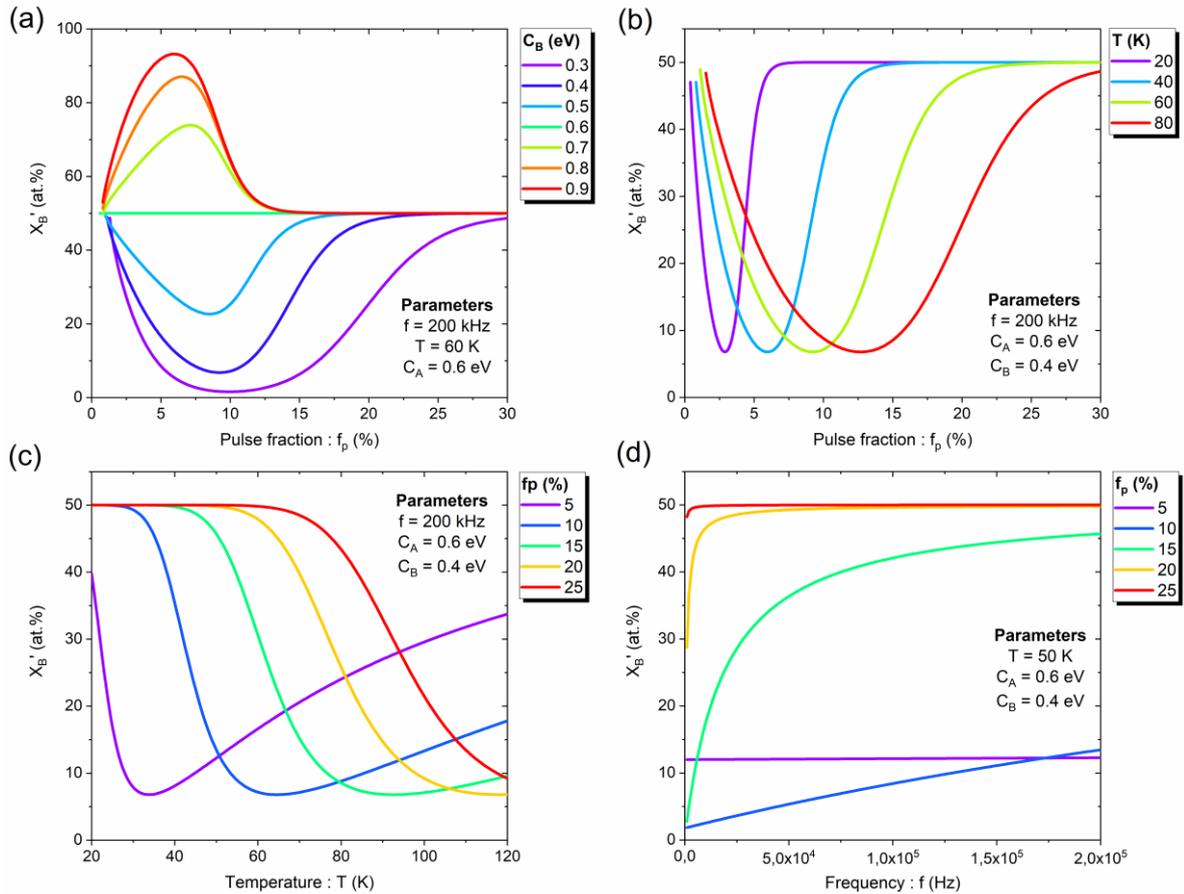

**Fig. 2**: Evolution of the apparent composition of B ($X'_B$ in at.%) for an equiatomic alloy (A/B) as a function of : (a) the pulse fraction ($f_p$ in %) for different B energetic constants ($C_B$ in eV), (b) also the pulse fraction but for different analysis temperatures ($T$ in K), (c) the temperature for different pulse fraction and (d) the pulse frequency ($f$ in Hz) for different pulse fractions

As expected, PE does not occur when the elements have the same energetic constant ($C_A = C_B$) (Fig. 2.a). Also as expected, PE is depending on whether the B atoms have higher energetic constant than A atoms ($C_A < C_B$ *i.e.* high field) or a lower one ($C_A > C_B$ *i.e.* low field) (Fig. 2.a). Regarding the different evolutions of the apparent composition as the function of the analysis conditions (Fig. 2), to reduce the PE of one element compared to the other (*i.e.* $X'_B$ close to $X_B$), it is required to increase pulse frequency ($f$) or the pulse contribution, $\alpha_p$ (and so the pulse fraction $f_p$), or by decreasing the base temperature. This behavior is well known experimentally, and analysis conditions must be tested to calibrate every new material (Lefebvre et al., 2016).

For any new material, the knowledge of the different energetic constants $C_i$ allows us to predict the apparent chemical composition as a function of the experimental parameters set by users ($T$, $f$ and $\alpha_p$), those depending on the APT used ($\Delta$) and the studied material ($X_i$). In this study, energetic constants values were determined for different alloys by fitting the predicted apparent compositions to the experimentally measured ones under different analysis conditions. The deviation between the calculated and experimental values was minimized using least square method.

## 3. APT analysis conditions and measurement procedure

### 3.1 Materials

Three different metallic alloys were analyzed by APT using a broad range of analysis conditions. The first alloy is composed of Ni and Cu in the same proportion (50 at.% of Ni and 50 at.% of Cu). This alloy has been chosen due to the significant difference of theoretical evaporation field between copper ($E_{Cu}$ = 30 V/nm) and nickel ($E_{Ni}$ = 35 V/nm) (Tsong, 2005). The second one is a Fe based metal composed of Cr and Ni with the following nominal chemical composition: 19.5 at.% of Cr and 11.7 at.% of Ni (Fe is in balance). This alloy is composed of 80% of austenite and 20% of ferrite. The austenite is enriched with Ni (12 at.% of Ni and 18 at.% of Cr) and the ferrite is enriched with Cr (10 at.% of Ni and 20 at.% of Cr), according to EDS measurements. The third alloy is also a Fe based metal with a nominal composition of 1.05 at.% of Cu. The choice of the last two alloys (FeCrNi and FeCu) was motivated by the large number of studies of Fe based metals utilizing APT. It remains some interrogations about the APT quantitative measurements of Cu composition in Fe matrix. This last case was chosen regarding the large number of former studies focused on the understanding on experimental APT biases. This case is historically a model case for preferential evaporation investigations (Takahashi et al., 2017; Takahashi & Kawakami, 2014; Shu et al., 2018).

### 3.2 APT analysis conditions

Three parameters are known to play a major role in the composition measurements accuracy by APT (Takahashi & Kawakami, 2014): the analysis temperature, the pulse repetition rate and the pulse fraction. In this study, a Cameca Local Electrode Atom Probe (LEAP) 4000 HR has been used for all the alloys. The influence of the sample temperature ($T$) has been studied in the range 20-90 K. The pulse repetition rate ($f$) influence has been studied in the range 25-250 kHz. The last parameter is the pulse fraction ($f_p$). $f_p$ influence has been studied in the range 2.5-30 %. For all the analyses, the detection rate was between 0.3 to 0.5%/pulse. Though we do not present the data, we also performed analyses varying only the detection rate (from 0.1 to 10%) to ensure that this parameter does not influence the apparent composition, as already observed by Takahashi.

For each set of analysis condition, an average of $0.6 \times 10^6$ impacts were collected. This value is sufficiently high to obtain a relatively low statistical uncertainty on the composition measurements (around $2\sigma = 0.14$ at.% for the NiCu system for example) and sufficiently low to avoid a significant morphological change of the sample during series of measurements (in particular the radius of curvature of the specimen).

### 3.3 Measurement procedure

In the NiCu case, Ni and Cu atoms are mainly evaporated as $Ni^{2+}$, $Cu^{2+}$ and $Cu^+$. There is a very small amount of $Ni^+$ ions (less than 0.7% of Ni ions). This can be well understood considering the Kingham post ionization theory (Kingham, 1982; Haydock & Kingham, 1980). The same behavior for FeCu is observed where Fe and Cu atoms are mainly evaporated as $Fe^{2+}$, $Cu^{2+}$ and $Cu^+$. In the case of FeCrNi, almost all atoms are evaporated as double charged ions (*i.e.* $Fe^{2+}$, $Cr^{2+}$ and $Ni^{2+}$).

It has to be noticed that O and N atoms were also detected for the NiCu alloy. Regarding the small amount (less than 0.3 at.%), only Ni and Cu atoms will be considered in the following. For the FeCu alloy, a small amount of C, P, O, Al, Cr, Mn and N atoms was detected (total amount less than 0.3 at.%). In this case, only Fe and Cu atoms will be considered. For the FeCrNi alloy, a small amount of C, O, N and Al atoms (less than 0.5 at.%) was also detected and similarly will not be taken into account in this study.

After identification of the peaks of interest in the mass spectra, the background into each peak was removed according to the background level from both sides of the peak as defined in (Lefebvre et al., 2016). To estimate the chemical composition, the overlaps of $^{54}Fe^{2+}$ with $^{54}Cr^{2+}$ and $^{58}Fe^{2+}$ with $^{58}Ni^{2+}$ were considered (FeCrNi alloy). This was done considering the natural abundance of isotopes. There is no isotopic overlap for the NiCu and the FeCu alloys.

The atoms located in the center of crystallographic poles were also not considered in the composition measurements. This region, which exhibits commonly a low atomic density, is prone to electrical field fluctuations, inducing chemical biases and significant detection bias due to the rapid evaporation of last atoms situated on a low index pole center. Moreover, the chemical bias is not the same from one crystallographic direction to another, as already been observed by Gault *et al.* (B. Gault et al., 2012).

If ions spatially and temporally give rise to close impacts the detector (so-called multi-hits events), an additional selected loss induced by detector loss of efficiency may occur and can induce chemical composition measurements biases (Meisenkothen et al., 2015). This composition bias was not considered in our model; therefore, a multi-hits correction was applied to experimental data following the procedure described in (Angseryd et al., 2011; Thuvander et al., 2011). Fig. 3 represents the results of such correction, applied to the FeCrNi alloy and the uncertainty corresponds to the standard deviation. The multi-hits correction magnitude is lowest at high temperature, as seen in Fig. 3. As also observed by field evaporation simulation (Gruber et al., 2011), detection issues, that could bias the apparent composition, are lower at higher temperature. So, considering only multi-hits detection issues, analyses at high temperature were preferred for quantitative chemical composition measurements.

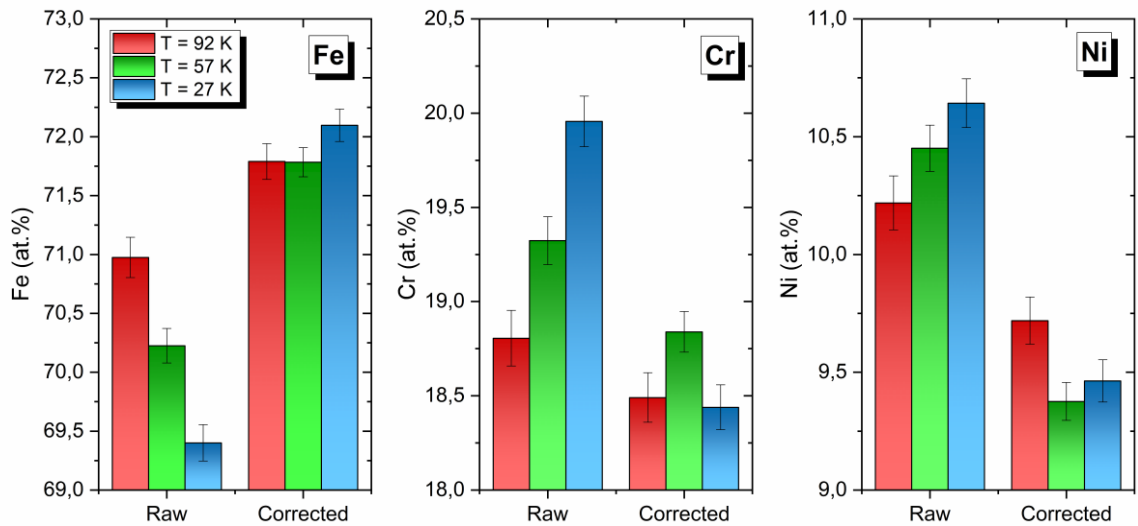

**Fig. 3**: Measured composition of Fe, Cr and Ni (at.%) in the FeCrNi alloy at high pulse fraction (25%) at 3 different temperatures (27, 57 and 92K) : composition measured from the raw data (Raw) and the one with multiple hits correction (Corrected).

## 4. Results

### 4.1 NiCu

The pulse frequency influence has been studied using two different temperatures and pulse fractions (fig.4.a). Whatever the conditions, the apparent composition of Cu decreases when decreasing the pulse frequency. In the studied range of pulse repetition rate, this trend is more pronounced at high temperature. The measured Cu composition decreases from 45.5 ± 0.2 at.% at 200 kHz pulse repetition rate to 27.0 ± 1.1 at.% at 25kHz at temperature of 92 K and a pulse fraction of 20 %. Whereas, at lower temperature (45 K) and a lower pulse fraction (10%), it decreases only from 45.9 ± 0.1 at.% to 42.1 ± 0.2 at.%. This influence of the pulse repetition rate can be easily interpreted in term of time interval between pulses. Indeed, as the pulse frequency increases, the duration during which PE can occur decreases, since the time interval between pulses becomes shorter. This explains why at high pulse repetition rate (200 kHz), the apparent composition is closer to the nominal (Fig. 2.d).

The influence of the pulse fraction has been also studied at two temperatures (92 and 57 K) for a constant repetition rate of 200 kHz (fig.4.b). Whatever the temperature, a similar trend is experimentally observed. The apparent composition of Cu decreases when the pulse fraction decreases. However, at a very low pulse fraction (5%) and high temperature, measured Cu amount goes up. At high pulse fractions, which means at low electric field between pulses, the evaporation probability of Ni and Cu atoms between pulses is relatively low, explaining that the apparent composition is close to the nominal. Reducing the pulse fraction means increasing the electrical field induced by the DC voltage, his tends to increase the evaporation probability between pulses. Since the evaporation field of Cu is lower than that of Ni, the probability to evaporate Cu between pulses is higher than Ni. This explains the PE of Cu and thus the decrease of the apparent Cu composition. By continuing to decrease the pulse fraction (and so increasing the electrical field at the DC voltage), the probability to evaporate Ni at the DC voltage increases significantly which balances the PE of Cu. It induces an increase of the apparent Cu composition (Fig. 2.b).

The estimation of the model parameters has been done considering that the exact Cu composition was 48 at.% (small deviation to the nominal 50at.% probably due to composition fluctuations in the material). The values of the constant $C_{Cu}$ and $C_{Ni}$ determined from the adjustment (Fig. 4) are not function of the temperature nor the electric field. The electric field has been estimated using the Kingham curves, based on the charge state ratio (Kingham, 1982; Haydock & Kingham, 1980), considering the work function associated to pure Cu. Whatever the experimental conditions are, the values of $C_{Cu}$ is lower than the value of $C_{Ni}$. The average value of $C_{Cu}$ is equal to 0.65 ± 0.04 eV and the one of $C_{Ni}$ is equal to 0.76 ± 0.06 eV. The uncertainty corresponds to the standard deviation.

(a)

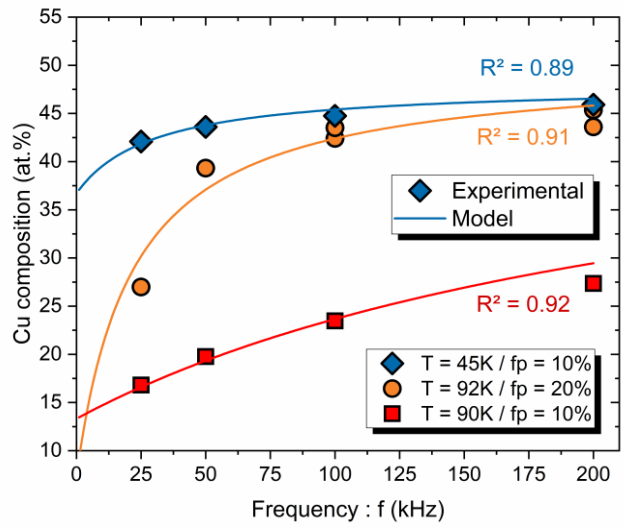

(b)

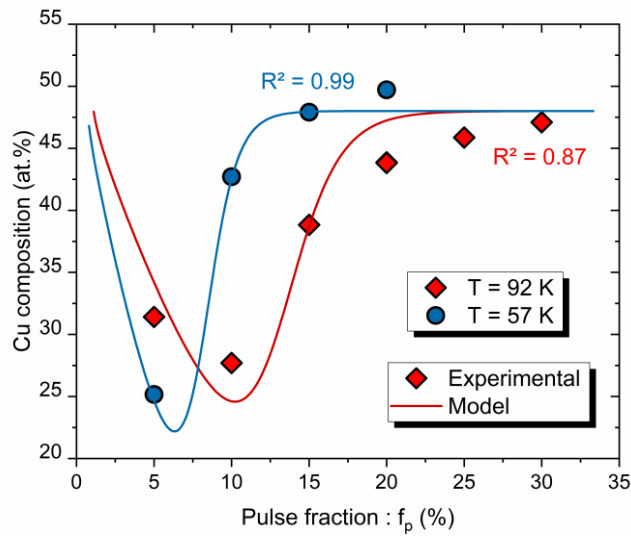

**Fig. 4:** Measured Cu composition (at.%) (Ni in balance) as a function of (a) the pulse frequency f (kHz) and (c) the pulse fraction $f_p$ (%) for different APT analysis conditions.

### 4.2 FeCrNi

Since the pulse fraction and the analysis temperature are the main parameters influencing the apparent composition, the study was focused on these 2 parameters (Fig. 2). The influence of the pulse fraction at three temperatures (27, 57 and 92 K) for a constant pulse frequency of 200 kHz is reported on Fig. 5.a and Fig. 5.b. Whatever the temperature, a similar trend is observed. The apparent Cr composition decreases when the pulse fraction decreases (Fig. 5.a), and inversely for the apparent Ni composition (Fig. 5.b). It should be highlighted that the apparent Fe composition is not constant and follows a similar trend like Ni. The model (Fig. 5.a and Fig. 5.b) reproduces well this trend. At high pulse fraction, the apparent compositions of Cr and Ni are close to the expected one. Reducing the pulse fraction down to a threshold value induces a decrease of the apparent Cr composition (and inversely for the Ni) since Cr is the element with the lowest evaporation field. Like in the NiCu alloy, evaporation at DC voltage becomes important at very low pulse fractions for all elements, whatever their energetic constant and their evaporation field are. As a consequence, the measured composition tends toward the real one. Since the FE probability is linked to the temperature, the pulse fractions for which PE of Cr is observed, and the pulse fraction for which all elements evaporate at DC voltage are higher at higher temperatures.

Conversely to the previous analysis (*i.e.* NiCu) where it was well known, here the composition depends on the studied phase (*i.e.* austenite or ferrite) and is only approximatively known from EDS measurements. Since the nominal composition is not perfectly known, the average one measured at high pulse fraction (> 10 %) was used. It corresponds to a nominal composition of Cr and Ni respectively equal to 18.6 and 9.5 at.%. These values are used to perform the adjustment of the analytical model (Fig. 5.a and Fig. 5.b, straight lines). In this case, the analytical model reproduced very well the experimental values and their associated evolution as a function of the analysis temperature and the pulse fraction.

Regarding the values of the energetic constant of Fe, Cr and Ni obtained from the analytical model, it is observed that : - (*i*) they are constant whatever the APT analysis conditions are (only 10% of deviation between extremum compared to the mean value) as for the NiCu case and - (*ii*) the highest constant of energy is for Ni and the lowest for Cr (like their respective theoretical evaporation fields (Tsong, 2005)). Since there is no influence of APT analysis conditions, the average values are: 1.13 ± 0.03 eV for Fe ($C_{Fe}$), 1.05 ± 0.03 eV for Cr ($C_{Cr}$) and 1.17 ± 0.03 eV for Ni ($C_{Ni}$). It appears that the average value of the energetic constant of Ni ($C_{Ni}$) is not the same in the case of the NiCu alloy ($C_{Ni}$ = 0.76 ± 0.04 eV) and in the case of FeCrNi steel ($C_{Ni}$ = 1.17 ± 0.03 eV). This point will be discussed later.

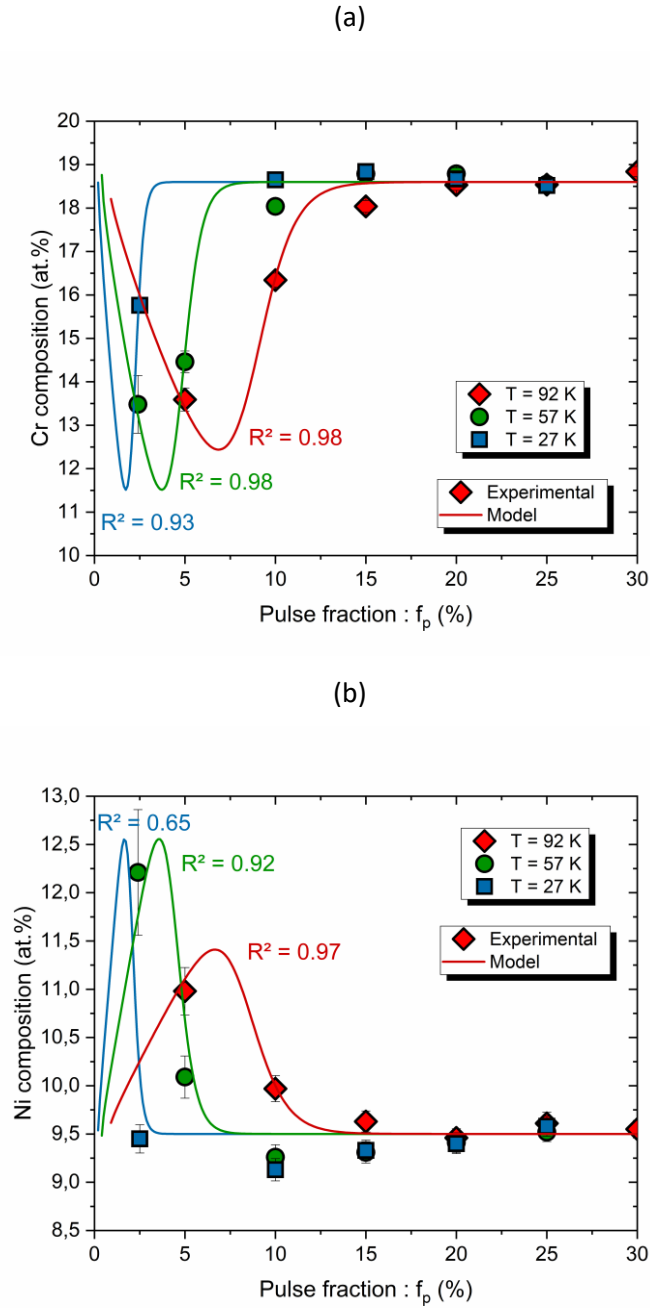

**Fig. 5**: Measured Cr (a) and Ni (b) compositions as a function of the pulse fraction ($f_p$ in %). Fe is in balance of the sum of these 2 graphs. Full lines are results of the analytical model.

### 4.3 FeCu

For the FeCu alloy, only the influence of the temperature at constant pulse fraction ($f_p$ = 20%) and pulse frequency (f = 200 kHz) was studied. As the temperature increases, the apparent Cu composition decreases (Fig. 6). The experimental trend is very well reproduced with the model (Fig. 6). Indeed, as the temperature increases, the evaporation probability increases, at the pulse and also at the steady voltage, inducing an increase of the PE probability (Fig. 2.c). So, it leads to a significant loss of Cu since it is the low field element.

The fit of the data also shows a slight increase of the apparent Cu composition for a temperature higher than 90 K. Indeed, continuing to increase the temperature, the evaporation probability of Fe between pulses may become significant so that it could balance the PE of Cu. This gives rise to an increase of the apparent Cu composition.

Regarding the values of the energetic constant of Fe and Cu obtained from the model, it is observed that the highest constant of energy is for Fe and the lowest is for Cu. The values are: $C_{Fe}$ = 0.38 ± 0.02 eV and $C_{Cu}$ = 0.34 ± 0.02 eV. The value of the energetic constants of Fe ($C_{Fe}$) in FeCu is not the same than in the FeCrNi alloy ($C_{Fe}$ = 1.13 ± 0.03 eV) and the value of the energetic constant of Cu ($C_{Cu}$) in FeCu is not the same than in the NiCu alloy ($C_{Cu}$ = 0.65 ± 0.03 eV).

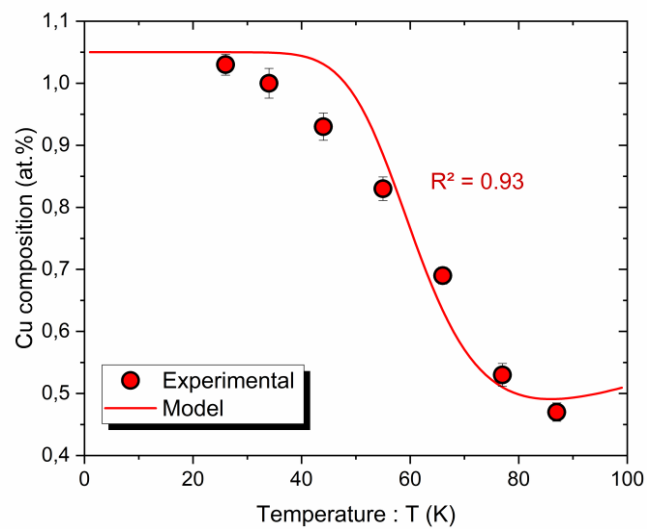

**Fig. 6**: Measured Cu composition as a function of temperature (K). In all cases, Fe is in balance and full line is the result of the analytical model. The pulse fraction and frequency are 20% and 200 kHz respectively.

## 5. Discussion

Regarding the results presented above, the analytical model reproduces relatively well all experimental measurements. The model reproduces the evolution of the apparent composition as a function of APT analysis conditions: temperature, pulse fraction and pulse repetition rate. This is already a big step forward as compared to the analytical model proposed by Takahashi *et al.* (Takahashi & Kawakami, 2014)

### 5.1 Optimal analysis conditions

As mentioned in the introduction, one of the main objectives of this study was to determine the optimal APT analysis conditions. For a specific alloy, at a defined temperature and pulse frequency, the pulse fraction from which there is no PE is the one from which the apparent composition is constant and is chosen as the optimal APT analysis conditions (Fig. 2.b). For example, for the NiCu alloy, at a temperature of 57 K and a pulse frequency of 200 kHz, the minimum pulse fraction to apply to avoid PE is close to 13 % (Fig. 4.b). With this approach, using the analytical model and the associated energetic constant obtained experimentally, it is possible for each alloy to determine the optimal APT analysis conditions to avoid PE. The results are shown in Fig. 7. For each alloy, the analysis temperature as a function of the pulse fraction at different pulse frequency is depicted. For example, in the NiCu alloy (Fig. 7.a), at a pulse fraction of 13% and a pulse frequency of 200 kHz (LEAP), the maximum analysis temperature to avoid PE is equal to 57 K (confirmed by our experimental). This data can be extrapolated to another APT.

Regarding the classical pulse fraction (*i.e.* 15 ≤ $f_p$ ≤ 25%) used by APT users, the analysis temperature must be lower than 80 - 90 K and 40 - 45 K for the NiCu (Fig. 7.a) and FeCu (Fig. 7.c) alloys respectively, to avoid PE. Whatever the temperature range is (*i.e.* from 20 to 100K), at this pulse fraction, there is no PE for the FeCrNi alloy during LEAP analysis (Fig. 7.b). The domain where there is no PE in the pulse fraction - temperature space, is very small for the FeCu alloy, indicating that it is very sensitive to PE.

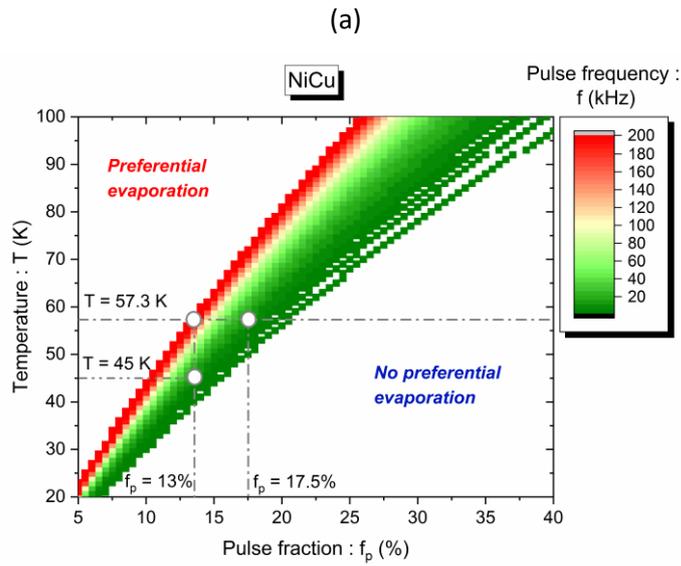

(a)

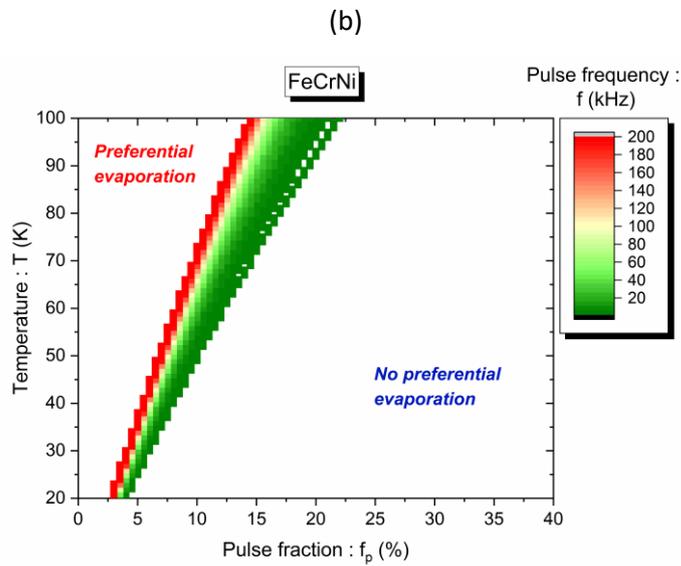

(b)

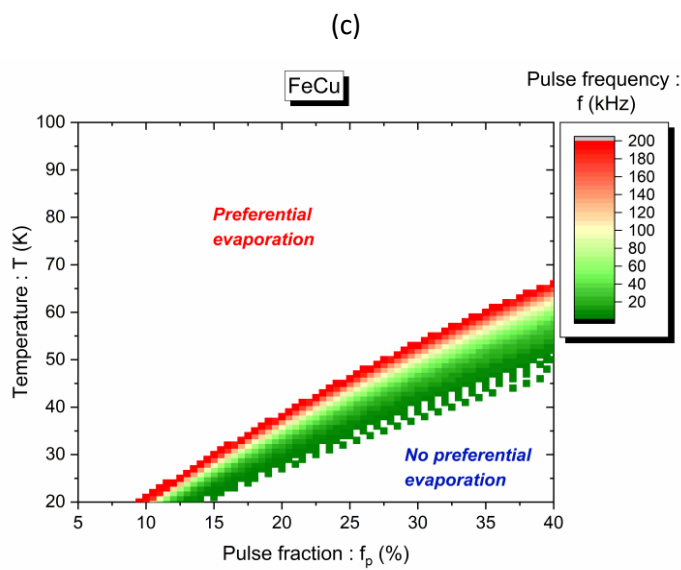

(c)

**Fig. 7**: Maximum analysis temperature (T in K) as a function of the pulse fraction ($f_p$ in %) at different pulse frequency (f in kHz) for which there is no PE, for different alloys: (a) NiCu, (b) FeCrNi and (c) FeCu.

**5.2 Energetic constant and extrapolation to other materials**

The average energetic constants ($C_i$) of Fe, Cr, Ni and Cu, obtained with the analytical model applied to different alloys are reported in the table 1.

**Table 1**: Energetic constant ($C_i$ in eV) of Fe, Cr, Ni and Cu as a function of the alloys of this study: NiCu, FeCrNi and FeCu. The uncertainty corresponds to the standard deviation.

|       | Fe          | Cr          | Ni          | Cu          |
|-------|-------------|-------------|-------------|-------------|
| NiCu  | -           | -           | 0.76 ± 0.06 | 0.65 ± 0.04 |
| FeCrNi| 1.13 ± 0.03 | 1.05 ± 0.03 | 1.17 ± 0.03 | -           |
| FeCu  | 0.38 ± 0.02 | -           | -           | 0.34 ± 0.02 |

All values obtained are in the order of eV (from 0.35 to 1.17 eV). Comparing these values with those of the literature is difficult because only few results are available. From data obtained by Ohnuma (Ohnuma, 2019) using DFT calculation, under intense electrostatic field, it can be roughly estimated that the energetic constant of Fe in FeCu alloy is lower than 2 eV. To the authors' knowledge, it is the only study related to the materials studied here. Using the same approach (DFT calculation, under intense electrostatic field), Carrasco *et al.* (Carrasco et al., 2018) and Peralta *et al.* (Peralta et al., 2013) have estimated that the constant energy of Al is respectively equal to 0.5 eV in pure Al and 1.1 eV in $Al_3Sc$ phase. These DFT results are very similar to our experimental results on several points: (*i*) the range of energetic constant (few tenth of eV) and (*ii*) the energetic constant is linked to the studied phase. For example, the energetic constant of Fe is equal to 1.13 eV in FeCrNi and equal to 0.38 eV in FeCu (Table 1). This can be easily understood since the energetic constant of an element is a fraction of its energy barrier without field ($Q_{0,i}$) (Kreuzer & Nath, 1987) which is linked to its binding-energy with the surrounding element (Gomer & Swanson, 1963; Ashton et al., 2020). The energetic constants must be determined before and for each phase in order to be able to estimate the optimal analysis conditions. These values of energetic constant can be obtained, as done in this study, by varying the APT analysis conditions and by then adjusting the apparent composition with the analytical model. Another possibility would be to obtain these values by DFT calculation. However, it must be first confirmed that DFT calculations provide quantitative energetic constant corresponding to those obtained experimentally. It must be noted that the present experimental work allows for the first time the measurement of these energetic constants for alloys. Indeed, to date, these values were only obtained experimentally on pure materials, such as W (Kellogg, 1984) or Rh (Ernst, 1979).

Since the energetic constants are tabulated values for a specific material, the values obtained in this study for the FeCu alloy (Fe-1.05Cu at.%) were applied to experimental measurements obtained by Takahashi *et al.* (Takahashi & Kawakami, 2014) on a very similar material (Fe-1.2Cu at.%). In their study, they measured the apparent composition of Cu as a function of the analysis temperature (35 – 100K) at different pulse fractions (15, 20 and 25%) using an energy-compensated 3DAP with a large-angle reflectron (Oxford Nanoscience Ltd.) at a pulse frequency of 20 kHz and a pulse width ($\Delta$) of 10 ns. Their measurements are reported in Fig. 8. Using the energetic constants obtained in this study for the Fe (0.38 eV) and Cu (0.34 eV, table 1) in the FeCu alloy does not reproduce the evolution of the apparent composition of Cu as a function of the APT analysis conditions (Fig. 8.a). In table 2, the values of the energetic constants of Fe and Cu obtained by fitting the experimental data of Takahashi with the analytical model of this study (Fig. 8.b) are reported. First of all, the values obtained are close to those of this study (Table 1). In addition, the values thus obtained are not constant as a function of the pulse fraction (Table 2) and tend towards the values of this study (Table 1) when the pulse fraction increases. This evolution probably results from the data treatment procedure of Takahashi. Indeed, in their study, they tried to develop an analytical model of the PE considering the noise level in the mass spectrum. Thus, the background noise in the apparent composition measurement was included, while it was corrected in our study. This explains why our values do not reproduce the experimental data of Takahashi (Fig. 8.a). As Takahashi also observed, when the pulse fraction increases, the noise level decreases drastically, explaining why the values obtained on the adjustment of these data tends towards those of this study at high pulse fraction, as the noise level is the lowest his influence on the measures is the lowest.

**Table 2**: APT analysis conditions (T, $f_p$, and f) applied to the Fe-1.2Cu (at.%) alloy (Takahashi & Kawakami, 2014). The values of $C_{Fe}$ and $C_{Cu}$ (eV) obtained from the adjustment of the experimental results by the analytical model developed in this study are also reported.

| T (K) | $f_p$ (%) | f (KHz) | $C_{Fe}$ (eV) | $C_{Cu}$ (eV) |
|---|---|---|---|---|
| 40 → 105 | 15 | 20 | 0.51 | 0.32 |
| 40 → 105 | 20 | 20 | 0.44 | 0.30 |
| 40 → 105 | 25 | 20 | 0.38 | 0.31 |

(a)

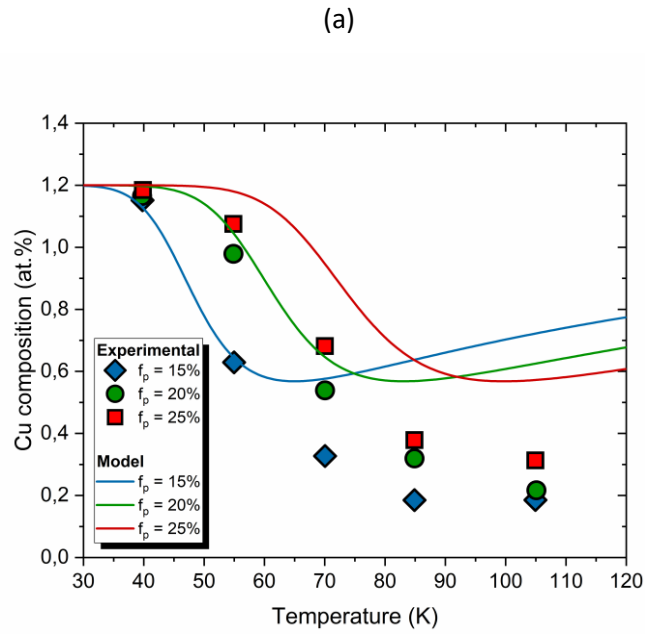

(b)

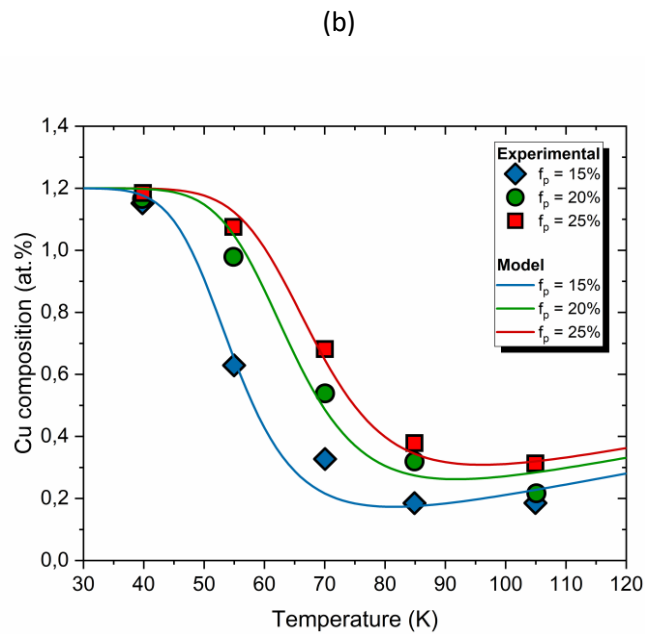

**Fig. 8**: Cu compositions (at.%) as a function of the analysis temperature (K) for different pulse fractions $f_p$ (15, 20 and 25%). Experimental data (dots) have been extracted from (Takahashi & Kawakami, 2014). Full lines are obtained from the analytical model of this study using energetic constants of Fe and Cu reported in (a) table 1 and (b) table 2.

## Conclusions

An analytical model is proposed that enables to quantify and predict the preferential evaporation. The validity of the model is supported by experimental measurements, on three different alloys (NiCu, FeCrNi and FeCu). The strong influence of the analysis temperature and of the pulse fraction on the apparent composition has been confirmed and to a lesser extent the influence of the pulse frequency. Those influences are well reproduced by the model. Thus, the model can estimate the best analysis conditions, avoiding PE and leading to quantitative composition measurements.

Even more, the model can also provide some valuable information about the energetic constant linked to the evaporation process. They are tabulated values for a defined phase. In this study, they have been obtained from experimental measurements. The next step will be to confirm the values of those energetic constants, with some numerical approach (such as DFT), and thus to be able to predict the best APT analysis conditions for any kind of phase.

## Acknowledgments

A part of this work has been performed on the Genesis platform. Genesis is supported by the Région Haute-Normandie, the Métropole Rouen Normandie, the CNRS via LABEX EMC3 and the French National Research Agency as a part of the program "Investissements d'avenir" with the reference ANR-11-EQPX-0020

## Bibliography

Amirifar, N., Lardé, R., Talbot, E., Pareige, P., Rigutti, L., Mancini, L., Houard, J., Castro, C., Sallet, V., Zehani, E., Hassani, S., Sartel, C., Ziani, A. & Portier, X. (2015). Quantitative analysis of doped/undoped ZnO nanomaterials using laser assisted atom probe tomography: Influence of the analysis parameters. *Journal of Applied Physics* **118**, 215703.

Angseryd, J., Liu, F., Andrén, H.-O., Gerstl, S. S. A. & Thuvander, M. (2011). Quantitative APT analysis of Ti(C,N). *Ultramicroscopy* **111**, 609–614.

Ashton, M., Mishra, A., Neugebauer, J. & Freysoldt, C. (2020). Ab initio Description of Bond-Breaking in Large Electric Fields. *arXiv:2002.02808 [cond-mat]*. http://arxiv.org/abs/2002.02808.

Bacchi, C., Costa, G. D. & Vurpillot, F. (2019). Spatial and Compositional Biases Introduced by Position Sensitive Detection Systems in APT: A Simulation Approach. *Microscopy and Microanalysis* **25**, 418–424.

Blum, I., Zanuttini, D., Rigutti, L., Vurpillot, F., Douady, J., Jacquet, E., Anglade, P.-M., Gervais, B., Vella, A. & Gaillard, A. (2016). Dissociation of Molecular Ions During the DC Field Evaporation ZnO in Atom Probe Tomography. *Microscopy and Microanalysis* **22**, 662–663.

Carrasco, T., Peralta, J., Loyola, C. & Broderick, S. R. (2018). Modeling field evaporation degradation of metallic surfaces by first principles calculations: A case study for Al, Au, Ag, and Pd. *Journal of Physics: Conference Series* **1043**, 012039.

Ernst, N. (1979). Experimental investigation on field evaporation of singly and doubly charged rhodium. *Surface Science* **87**, 469–482.

Forbes, R. G. (1995). Field evaporation theory: a review of basic ideas. *Applied Surface Science* **87–88**, 1–11.

Gault, B., Danoix, F., Hoummada, K., Mangelinck, D. & Leitner, H. (2012). Impact of directional walk on atom probe microanalysis. *Ultramicroscopy* **113**, 182–191.

Gault, Baptiste, Moody, M. P., Cairney, J. M. & Ringer, S. P. (2012). *Atom Probe Microscopy*. 2012th ed. Springer.

Gomer, R. & Swanson, L. W. (1963). Theory of Field Desorption. *The Journal of Chemical Physics* **38**, 1613–1629.

Gruber, M., Vurpillot, F., Bostel, A. & Deconihout, B. (2011). Field evaporation: A kinetic Monte Carlo approach on the influence of temperature. *Surface Science* **605**, 2025–2031.

Hatzoglou, C., Radiguet, B. & Pareige, P. (2017). Experimental artefacts occurring during atom probe tomography analysis of oxide nanoparticles in metallic matrix: Quantification and correction. *Journal of Nuclear Materials* **492**, 279–291.

Haydock, R. & Kingham, D. R. (1980). Post-Ionization of Field-Evaporated Ions. *Physical Review Letters* **44**, 1520–1523.

Kellogg, G. L. (1984). Measurement of activation energies for field evaporation of tungsten ions as a function of electric field. *Physical Review B* **29**, 4304–4312.

Kingham, D. R. (1982). The post-ionization of field evaporated ions: A theoretical explanation of multiple charge states. *Surface Science* **116**, 273–301.

Kreuzer, H. J. & Nath, K. (1987). Field evaporation. *Surface Science* **183**, 591–608.

Lefebvre, W., Vurpillot, F. & Sauvage, X. (2016). *Atom probe tomography: put theory into practice*. Boston, MA: Elsevier.

Meisenkothen, F., Steel, E. B., Prosa, T. J., Henry, K. T. & Prakash Kolli, R. (2015). Effects of detector dead-time on quantitative analyses involving boron and multi-hit detection events in atom probe tomography. *Ultramicroscopy* **159**, 101–111.

Miller, M. K. (2000). *Atom Probe Tomography: Analysis at the Atomic Level*. Springer US.

Miller, M. K., Cerezo, A., Hetherington, M. G. & Smith, G. D. W. (1996). *Atom probe Field Ion Microscopy*.

Miller, M. K. & Forbes, R. G. (2009). Atom probe tomography. *Materials Characterization* **60**, 461–469.

Muller, E., Nakamura, S., Nishikaw.o & McLane, S. (1965). Gas-Surface Interactions and Field-Ion Microscopy of Nonrefractory Metals. *Journal of Applied Physics* **36**, 2496-.

Ohnuma, T. (2019). Surface Diffusion of Fe and Cu on Fe (001) Under Electric Field Using First-Principles Calculations. *Microscopy and Microanalysis* **25**, 547–553.

Peng, Z., Choi, P.-P., Gault, B. & Raabe, D. (2017). Evaluation of Analysis Conditions for Laser-Pulsed Atom Probe Tomography: Example of Cemented Tungsten Carbide. *Microscopy and Microanalysis* **23**, 431–442.

Peralta, J., Broderick, S. R. & Rajan, K. (2013). Mapping energetics of atom probe evaporation events through first principles calculations. *Ultramicroscopy* **132**, 143–151.

Prosa, T. J., Strennen, S., Olson, D., Lawrence, D. & Larson, D. J. (2019). A Study of Parameters Affecting Atom Probe Tomography Specimen Survivability. *Microscopy and Microanalysis* **25**, 425–437.

Russo, E. D., Blum, I., Houard, J., Costa, G. D., Blavette, D. & Rigutti, L. (2017). Field-Dependent Measurement of GaAs Composition by Atom Probe Tomography. *Microscopy and Microanalysis* **23**, 1067–1075.

Sánchez, C. G., Lozovoi, A. Y. & Alavi, A. (2004). Field-evaporation from first-principles. *Molecular Physics* **102**, 1045–1055.

Saxey, D. W. (2011). Correlated ion analysis and the interpretation of atom probe mass spectra. *Ultramicroscopy* **111**, 473–479.

Shu, S., Wirth, B. D., Wells, P. B., Morgan, D. D. & Odette, G. R. (2018). Multi-technique characterization of the precipitates in thermally aged and neutron irradiated Fe-Cu and Fe-Cu-Mn model alloys: Atom probe tomography reconstruction implications. *Acta Materialia* **146**, 237–252.

Takahashi, J. & Kawakami, K. (2014). A quantitative model of preferential evaporation and retention for atom probe tomography. *Surface and Interface Analysis* **46**, 535–543.

Takahashi, J., Kawakami, K. & Raabe, D. (2017). Comparison of the quantitative analysis performance between pulsed voltage atom probe and pulsed laser atom probe. *Ultramicroscopy* **175**, 105–110.

Thuvander, M., Weidow, J., Angseryd, J., Falk, L. K. L., Liu, F., Sonestedt, M., Stiller, K. & Andrén, H.-O. (2011). Quantitative atom probe analysis of carbides. *Ultramicroscopy* **111**, 604–608.

Tsong, T. T. (2005). *Atom-Probe Field Ion Microscopy: Field Ion Emission, and Surfaces and Interfaces at Atomic Resolution*. Cambridge University Press.

Wada, M. (1984). On the thermally activated field evaporation of surface atoms. *Surface Science* **145**, 451–465.

Zanuttini, D., Blum, I., Rigutti, L., Vurpillot, F., Douady, J., Jacquet, E., Anglade, P.-M. & Gervais, B. (2017). Simulation of field-induced molecular dissociation in atom-probe tomography: Identification of a neutral emission channel. *Physical Review A* **95**, 061401.